\documentclass[fleqn,twoside]{article}
\usepackage[headings]{espcrc2}

\readRCS
$Id: espcrc2.tex,v 1.2 2004/02/24 11:22:11 spepping Exp $
\usepackage{epsfig}
\usepackage[figuresright]{rotating}

\newcommand{\AmS}{{\protect\the\textfont2
  A\kern-.1667em\lower.5ex\hbox{M}\kern-.125emS}}

\newcommand{\p}{^{\prime}}
\def\be{\begin{equation}}
\def\ee{\end{equation}}
\def\eea{\end{eqnarray}}
\def\bea{\begin{eqnarray}}
\def\ep{\varepsilon}
\def\dd{\partial}
\def\o{\omega}
\def\k{\kappa}
\def\half{\frac{1}{2}}
\def\as{\bar \alpha_S}

\title{ Resummation of small $x$ contributions
to hard-scattering amplitudes}

\author{ R. Kirschner\address[ITPL]{Institut f\"ur Theoretische
Physik, Universit\"at Leipzig, \\
PF 100 920, D-04009 Leipzig, Germany}
        and
        M. Segond\addressmark[ITPL]\thanks{e-mail:roland.kirschner@itp.uni-leipzig.de;
mathieu.segond@itp.uni-leipzig.de} }

\runtitle{Small $x$ resummation}
\runauthor{R. Kirschner}

\begin{document}

\begin{abstract}
The summation of the small $x$ corrections to hard
scattering QCD amplitudes by collinear factorisation
method is reconsidered and the K-factor is derived in
leading $\ln x$ approximation. The corresponding
expression by Catani and Hautmann (1994) has to be corrected. 
The significance of the correction
is demonstrated in the examples of structure function
$F_L$ and of exclusive electroproduction.
\vspace{1pc}
\end{abstract}

\maketitle

\section{Small $x$ resummation }

Semi-hard processes are characterized by two essentially different large
momentum scales, the hard-scattering scale $Q^2$ and the large c.m.s. energy
squared $s$, $x$ being the small ratio of theses scales.  
The QCD calculation of the hard
processes involving the factorization of collinear singularities has to be
improved by including  the corrections enhanced by the large logarithm of
$x$. The results of the  QCD Regge asymptotics 
\cite{BFKL} provide the basis
for the resummation  of these large corrections. The method of fitting the
BFKL solution consistently into the collinear factorization, called also 
$k_T$ factorization, has been developed by M. Ciafaloni and collaborators
starting in 1990 \cite{CCH,CH,CC}. 

The resummed  small $x$ corrections affect the hard-scale evolution of the
parton distributions in terms of the anomalous dimension of two-gluon
composite operators and generate a K-factor that can be viewed as an
improvement of the coefficient function.  
Quite a number of papers is relying on this scheme
in general and on the results given in \cite{CH} 
in particular, e.g 
\cite{ASt,ABF,Ball:2001pq,Marzani:2008uh,Marzani:2008az,Diana:2009xv}.  

Small $x$ resummation was of great importance for  physics at 
HERA and it will be even more important for 
LHC physics. Motivated by this and by the idea of 
extending the applications to exclusive semi-hard 
production \cite{Ivanov:2007je}
 we have reconsidered the small $x$ resummation.
We confirm the known factorization scheme 
but disagree with
the expression for the K-factor given in \cite{CH} and approved in
\cite{CC}. Unlike the latter our expression has the angular momentum 
singularity  of the BFKL solution. In examples of the structure function
$F_L$ and vector meson electroproduction we demonstrate the significance
of the discrepancy. Clearly, the K-factor has to be corrected on the
leading log level before advancing to the next-to-leading log level.


\subsection{Scheme of resummation}

The amplitude of a hard scattering process calculated in the 
collinear factorization scheme has the structure of the convolution of
the coefficient function with the (generalized) parton distribution, 
\be
 A = C_A^{(0)} \otimes GPD, 
\ee
where the hard-scale dependence of the GPD is calculated from
the DGLAP/ERBL  equation \cite{DGLAP,ERBL}, 
summing the contributions enhanced by logs
of the hard scale $Q^2$. At large energy squared $s$ or small 
$x \approx \frac{Q^2}{s}$ the gluon contribution dominates; the
corresponding anomalous dimension is denoted by $\gamma_{\o}^{(0)}$.
In the calculation the parton distribution is emerging formally from a
bare distribution by absorbing the factorized collinear singularities.

At small $x$ corrections enhanced by logs of $x$ have to be summed,
improving both the coefficient function and the evolution kernel.
This resummation is done relying on the BFKL approach, where the
considered amplitude is calculated from the convolution of the
two reggeized-gluon Green function $g$ with impact factors $\Phi$,
\be
 A_{BFKL} = \Phi_A \otimes g \otimes \Phi_B. 
\ee
The consistent small $x$ improvement of the 
collinear factorization result
(1) requires the identification of  collinear singularities in
the BFKL amplitude and their factorization according to the adopted
scheme (e.g.$ \overline {MS}$). The corrections improve the evolution
kernel or the anomalous dimensions, 
$\gamma_{\o}^{(0)} \to \gamma_{\o}$,
and result in a correction factor $R_{\o}$ 
which appears as the
improvement of the coefficient function, 
$C_A (\o) = C_A^{(0)}(\o) R_{\o}$.

Actually there are no collinear singularities 
in the BFKL amplitude (2)
as long as the impact factors refer to the scattering of colourless
particles. It takes to replace $\Phi_B$ by a partonic impact factor
in convolution with a (bare) parton distribution. The projection of this
partonic impact factor onto the channel isotropic in the azimuthal angle
is simply constant in transverse momentum and its convolution with the
BFKL Green function $g$ results in the collinear singularities which are factorized to all order of the strong coupling constant  into an universal (but renormalization scheme dependent) transition function $\Gamma$, according to 
\be
F^{(0)}= g\otimes \Phi^{part} 
= F \cdot \Gamma , 
\ee
where the K-factor $R_{\o}$ appears in 
\be
 F( \o,\k,\mu_F) = \gamma_{\o} \; R_{\o} 
\left ( \frac {\k^2}{\mu_F^2} \right )^{\gamma_{\o}} \;
\ee
and then improves the coefficient function as seen above, whereas
the  transition function $\Gamma $ will  be absorbed into the renormalized (generalized) parton distribution.

\subsection{BFKL and collinear singularities}

Consider the BFKL equation in the forward limit in $2+2\ep$
dimensions,
\be
\o g(\o,\vec \k, \vec \k^0) =  
\delta^{2+2\ep} (\vec \k - \vec \k_0) + 
\bar \alpha_S \hat K \cdot 
g (\o, \vec \k, \vec \k_0 )
\ee
In one-loop approximation the operator $\hat K$ 
acts as 
$$ \hat K \cdot g (\vec \k, \vec \k_0 ) = 
\frac{1}{\pi} 
\int \frac{d^{2+2\ep}\k\p }{(2 \pi)^\ep}  
<\vec \k | \hat K |\vec \k\p > 
g (\vec \k\p, \vec \k_0 ),$$
where 
\be
<\vec \k | \hat K |\vec \k\p > = 
\frac{1}{(\vec \k - \vec \k\p)^2}  -
 \delta (\vec \k - \vec \k\p )
\alpha_g (\k), 
\ee
and the gluon trajectory reads
\be
\alpha_g (\k) = \half
\int \frac{d^{2+2\ep} \k^{\prime \prime} \vec \k^2}
{\vec \k^{\prime \prime \ 2} 
(\vec \k - \vec \k^{\prime \prime} )^2 }.
\ee
We have introduced $ \bar \alpha_S = \frac{ g^2 N_C}{ 8 \pi^2} $. 
For simplicity we restrict ourselves here to the exchange channel
$n=0$ represented by functions invariant under transverse rotations. 

 We  calculate 
the action of the kernel on functions 
that would be eigenfunctions at $\ep=0$.
\be
\hat K \cdot  (\vec \k^2)^{\gamma-1}
 = \lambda (\gamma, \ep) 
(\vec \k^2)^{\gamma -1 + \ep}, 
\ee
\be
\lambda (\gamma, \ep) = \frac{1}{(4 \pi)^{ \ep}}  
\left [ b(\gamma,  \ep) - \half b(0, \ep) 
\right ],
\ee 
\be
b(\gamma,  \ep) = \Gamma^{-1 } (\ep)
B(\ep, 1-\gamma-\ep) \ 
B(\ep, \gamma + \ep).
\ee
The solution for the Green function in the $n=0$ channel
can be formulated using the quasi-eigenvalues $\lambda (\gamma, \ep)$
and the operator of shifts in $\gamma$ by $\ep$,
\bea
g_0 (\k,\k_0) &=&  
\int_{-i\infty}^{i\infty} \frac{d \gamma}{2\pi i} \;  (\k^2 )^{\gamma }  
\\
&&\times      \frac{1 }
{ \o - \bar \alpha_S \  e^{-\ep \dd_{\gamma} } 
\lambda (\gamma,  \ep) }\; (\k_0^2)^{-\gamma } \;. \nonumber
\eea
Now $g_0$ is convoluted with the partonic impact factor:
\be
F^{(0)}(\o, \k,\ep)=\int \frac{d^{2 +2\ep} \k_0 }{\k_0^2 } 
 g_0 (\k, \k_0, \ep ) \  \Phi^{part} 
\ee
with $\Phi^{part}  = 
\frac{\bar\alpha_S }{ \mu^{2\ep}}
  \Theta (\mu_F^2- \k_0^2 )$, $\mu$ being the dimensional regularization scale.
The singularity in the integration over $\k_0$ is regularized 
in the infrared by $\ep$ with a positive real part and 
the cut-off at the factorization scale $\mu_F$ is unavoidable to 
prevent the divergence at large $\k_0$. 
\be
F^{(0)}(\o, \k,\ep)
=
\bar \alpha_S (\frac{\mu_F^2}{\mu^2})^{\ep}
\int \frac{d \gamma}{2\pi i} (\frac{\k^2}{\mu_F^2} )^{\gamma}
\ee
$$
\times \frac{1   }{ \o - \bar \alpha_S  
\frac{1}{\gamma} \lambda_1 (\gamma,\ep) 
e^{-\ep \dd_{\gamma} } }
\frac{1}{\ep - \gamma}
$$
The notation  $\lambda_1 (\gamma,\ep) = \gamma \lambda (\gamma- \ep,\ep)$ 
is introduced. 
In order to understand how the asymptotics $\ep \to 0$ is extracted it is
instructive to consider first the case where $\lambda_1 (\gamma,\ep)$ is
substituted simply by 1, which corresponds to the double-log approximation.
The shift operator can be treated easily after decomposition into geometric series
and this results without further approximations in
\be
F^{(0)}_{d.l.}
 =  \frac{\bar \alpha_S}{\o} (\frac{\mu_F^2}{\mu^2})^{\ep}
(\frac{\k^2}{\mu_F^2 })^{ \gamma_{\o}^{(0)}}   
\left ( \exp (\frac{1}{\ep} \gamma_{\o}^{(0)} ) - 1  
\right ).
\ee
In this crude approximation we obtain $R_{\o}= 1$ and also 
no corrections to the anomalous dimension $\gamma_{\o}^{(0)}$.
 The case without the simplification in 
$\lambda_1 (\gamma,\ep)$ gets closer to the considered one by the
substitution 
$ \tilde \gamma = \frac{\gamma}
{\lambda_1 (\gamma, \ep) } $.
We obtain
\bea
\nonumber
F^{(0)} (\o, \k,\ep)&=& 
\frac{\bar\alpha_S }{\o } (\frac{\mu_F^2}{\mu^2})^{\ep} \int 
\frac {d \tilde \gamma}{2\pi i}
\frac{\lambda_1 (\gamma, \ep)}{1- \tilde \gamma   
\lambda\p_1 (\gamma, \ep)} \\
 && \hspace{-1.0cm} \times (\frac{\k^2}{\mu_F} )^{\gamma}   
\frac{1}{ \tilde \gamma -  \hat \gamma_{\o}^{(0)}}
 \tilde \gamma \ \frac{1}{\ep -\gamma }  
\eea
where we have defined the following operator 
\be
\hat \gamma_{\o}^{(0)}= \frac{\bar\alpha_S }{\o } e^{-\ep \dd_{\gamma} } \;.
\ee
The factor arising by this change of integration 
variable involves
$\lambda_1\p (\gamma, \ep) =
\dd_{\gamma} \lambda_1 (\gamma, \ep)$
and it  becomes the essential
ingredient in the result for the K-factor $R_{\o}$. 
Now the operator ordering takes more care and 
results in
\be
\label{premresult}
F^{(0)}(\o,\k,\ep) =  
\frac{\bar\alpha_S}{\o} 
\frac{1}{1-  \gamma_{\o}^{(0)} \lambda\p_1 ( \gamma_{\o}, 0) }   
(\frac{\k^2}{\mu_F^2} )^{ \gamma_{\o} }
\ee 
$$   
\times  \exp \left (\frac{1}{\ep} \int_0^1 
\frac{d\alpha}{\alpha} \gamma_{\o} (\gamma_{\o}^{(0)} \alpha, \ep ) 
\right ) \; .
$$
For simplicity we have absorbed the $(\mu_F^2/\mu^2)^{\ep}$ factor in $\bar\alpha_S$, which is  from now also hidden in $  \gamma_{\o}^{(0)} = \frac{\bar \alpha_S}{\o}$. Note that this factor is in agreement with the renormalization group requirement for the coupling constant in a dimensional regularization. \\
The gluon anomalous dimension $\gamma_{\o}(\gamma_{\o}^{(0)}, \ep)$
with corrections resummed is the solution of 
the equation
\be 
\gamma_{\o} = \gamma_{\o}^{(0)} \ 
\lambda_1 (\gamma_{\o}, \ep). 
\ee
We expand the solution  in $\ep $,
\be
\gamma_{\o}(\gamma_{\o}^{(0)}, \ep)=
\gamma_{\o}+ \ep \, \gamma_{\o}\p + ...
\ee
From this equation we obtain $\gamma_{\o}$ 
by iteration
 as a power series in  $ \gamma_{\o}^{(0)}$
 and we show the first terms of the well known 
result of the
all order expansion of the BFKL anomalous 
dimension 
\begin{eqnarray}
\label{BFKLanomdim}
 \gamma_{\o}= \gamma_{\o}^{(0)} + 2 \zeta(3) \left( \gamma_{\o}^{(0)} \right)^4 \hspace{-0.1cm}+ 2 \zeta(5) \left( \gamma_{\o}^{(0)} \right)^6 \hspace{-0.1cm}+...
\end{eqnarray} 
and also of its $\ep$-correction term
\begin{eqnarray}
 \gamma_{\o}\p=  2 \zeta(3) \left( \gamma_{\o}^{(0)} \right)^3 \hspace{-0.1cm}-3 \zeta(4) \left( \gamma_{\o}^{(0)} \right)^4 \hspace{-0.1cm}+...
\end{eqnarray}

\subsection{K-factor}
We can now write the final result as 
\be
\label{finalresult}
F^{(0)} (\o, \k,\ep) = F(\o, \k,\mu_F) \; \Gamma \left(\gamma_{\o}^{(0)} (\frac{\mu_F^2}{\mu^2})^{\ep} , \ep \right) 
\ee
where 
\be
 F( \o,\k,\mu_F) = \gamma_{\o} \;R_{\o} 
\left ( \frac {\k^2}{\mu_F^2} \right )^{\gamma_{\o}} \;.
\ee
We rewrite the preexponential factor of (\ref{premresult}) in terms 
of the standard notation $\chi(\gamma) = 2 \psi(1) - \psi(1-\gamma) -
\psi(\gamma)$ for the BFKL eigenvalue function (at $\ep = 0, n=0$)
and its derivative.  The  collinear singularities, appearing from the dimensional regularization analysis  as a series of pole in $1/\ep$, are now  factorized into the
$\overline {MS}$ scheme gluon transition function 
\be
\Gamma (\gamma_{\o}^{(0)} , \ep) =   
\exp \left (\frac{1}{\ep} \int_0^{S_{\ep}}  
\frac{d\alpha}{\alpha} 
\gamma_{\o} ( \gamma_{\o}^{(0)}  \alpha ) \right )
\ee
with   the  corresponding usual factor 
$S_{\ep} = \exp \{ -\ep [ \psi (1) + \ln 4 \pi ]\} $.
 In eq.(\ref{finalresult}), we have written 
explicitly the factor 
$(\frac{\mu_F^2}{\mu^2})^{\ep}$ in front of 
$\gamma_{\o}^{(0)}$   to make clear that 
$F^{(0)} $ is $\mu_F$-independent, as we can see 
from the previous expression of $\Gamma$. 
We obtain  the K-factor as
\be
\label{Kfactor}
 R_{\o}(\gamma_{\o}^{(0)})=    \frac{1}
{ -  \gamma_{\o}^2 \chi\p(\gamma_{\o}) }  
  \ e^{L (\gamma_{\o})},
\ee $$ 
L (\gamma_{\o})\hspace{-0.05cm}  =\hspace{-0.1cm} \frac{1}{2}
\int_0^{\gamma_{\o}} \hspace{-0.4cm}
d\gamma \hspace{0.05cm} {2 \psi'(1)-\psi'(1-\gamma)-\psi'(\gamma) 
\over
  \chi(\gamma)}+\chi(\gamma).
$$
In this form the result can be compared with the one
by Catani and Hautmann \cite{CH,CC}. There the
exponential factor is the same whereas the preexponential 
one is the square root of ours. We plot these K-factors as a function  of $\gamma $ in the physical range $[0,1/2]$:

\begin{figure}[htbp]
\begin{center}
\epsfig{file=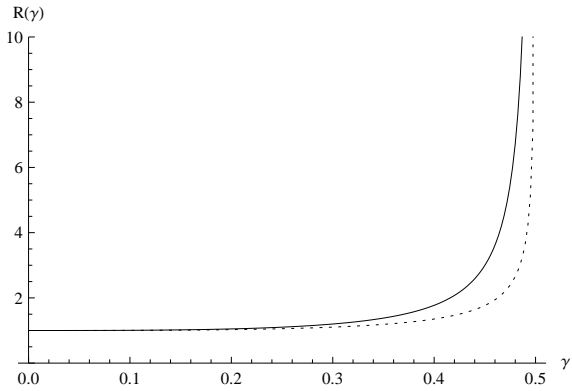,width=7.5cm}
\end{center}
\caption{K-factor $R(\gamma)$, our result (solid)  
and Catani-Hautmann's result (dashed)}
\end{figure}

Our K-factor is clearly enhanced  in comparison 
to the one of Catani-Hautmann, 
in particular when going towards the saturation 
value of the anomalous dimension.\\
 With our preexponential the dependence on $\o$ 
has the square root branch point characteristic 
for the BFKL solutions. 

We have checked that the equation used in \cite{CH}
is equivalent to  BFKL without modifications
and therefore the singularity structure of the
solution has to be the one of standard BFKL. 
 The solution given primarily in that paper 
in terms of sum can be shown to lead to our result.
The discrepancy is caused by 
restricting to a distinguished anomalous dimension 
in solving BFKL equation.  In \cite{CC} the 
argument involves an incorrectness in the analysis
of the $\ep \to 0$ asymptotics.

\section{Structure function $F_L$}
As an application of the previous discussion, we can  write the factorized expression of the longitudinal structure function $F_L$ in the Mellin space, in agreement with \cite{CH}

\begin{equation}
\label{FL}
F_L^{\o} (Q^{2})=C_{L , \, {\o}}^g(\as , Q^{2}/\mu^{2}_{F})  \;\; f^g_{\o}(\mu^{2}_{F}) \;,
\end{equation}
where $f^g_{\o}(\mu^{2})$ is the ${\o}-$moment of the renormalized gluon distribution function and we have defined  the gluonic (improved) coefficient function 

\begin{equation}
C_{L , \, {\o}}^g(\as , Q^{2}/\mu^{2}_{F}) =
h_{L , \, {\o}} \left( \gamma_{{\o}} \right) \;
R_{{\o}} \;(Q^{2}/\mu^{2}_{F})^{\gamma_{{\o}}} 
\label{CLNg}
\end{equation}
with the Mellin transform of the corresponding hard cross-section 

\begin{equation}
\label{hL}
 \, h_{L, \, {\o} } ( \gamma_{\o}) = {  \bar \alpha_S \over { 2 \, \pi}} \, N_f \, T_R \,
{ { 4 \, ( 1 - \gamma_{\o}) } \over { 3 - 2  \gamma_{\o} }} \, { {
\Gamma^3 ( 1 - \gamma_{\o})} \over {
\Gamma ( 2 - 2 \gamma_{\o}) } } 
\;\;
\end{equation}
$$ \times   { { \Gamma^3 ( 1 + \gamma_{\o}) } \over {
 \Gamma ( 2 + 2 \gamma_{\o}) } } \, . $$
Then we  easily obtain from (\ref{Kfactor}) and 
(\ref{BFKLanomdim}) the all order  
perturbative expansion in power of  
$\gamma_{\o}^{(0)}$ of this coefficient 
function for the simpler case $\mu^{2}_{F}=Q^2$
\label{CMSL}\nonumber
$${C_{L , \, {\o}}^{g}}(\as ,  1 )
= \frac{\as}{2\pi} T_R N_f { 4 \over 3}
\left\{\hspace{-0.05cm} 1 -\frac{1}{3} \gamma_{\o}^{(0)}\hspace{-0.05cm} +\hspace{-0.05cm}
\left[\hspace{-0.05cm}\frac{34}{9}\hspace{-0.05cm}-  \hspace{-0.05cm} \zeta (2)  \hspace{-0.05cm} \right]  \right. 
$$
$$   \times \left(\gamma_{\o}^{(0)} \right)^2 +
 \left[ {1 \over 3} \, \zeta (2) - \frac{40}{27} 
+
{14 \over 3} \, \zeta (3)    \right] \,
 \left(\gamma_{\o}^{(0)} \right)^3 $$
 $$+
 \left. \left[  \frac{1216}{81}-
{34 \over 9} \, \zeta (2) -
{20 \over 9} \, \zeta (3)  - 6 \, \zeta (4)  \right] \,
 \left(\gamma_{\o}^{(0)} \right)^4 \right. $$
 \begin{equation}
 +  \left.  {\cal O}\left(\left(\gamma_{\o}^{(0)} \right)^5\right) \right\}
 \end{equation}
\nonumber
$$\hspace{-1.0cm}  \simeq \frac{\as}{2\pi} T_R N_f \frac{4}{3} \hspace{-0.1cm} \left\{1\hspace{-0.05cm}-\hspace{-0.05cm} 0.33  \gamma_{\o}^{(0)} \hspace{-0.05cm}+\hspace{-0.05cm}
2.13 \left(\gamma_{\o}^{(0)} \right)^2 \hspace{-0.05cm}
  \right. $$
\bea
\nonumber
\label{numCFL}
\left. + \, 4.68 \left(\gamma_{\o}^{(0)} \right)^3  \hspace{-0.05cm}-0.37 \left(\gamma_{\o}^{(0)} \right)^4  \hspace{-0.1cm} + {\cal O}\left(\left(\gamma_{\o}^{(0)} \right)^5\right) \right\} \nonumber
\eea
which has to be compared with eq.(5.24) of 
Catani-Hautmann's paper \cite{CH}. 
Indeed, the results start to deviate 
 at the fourth loop in the  perturbative
expansion. 
The calculation made by 
Moch, Vermaseren and Vogt \cite{Moch}
on the same coefficient function derived by complete 
loop calculations in pure collinear factorization 
scheme extends to three loops and is, therefore, 
still not sufficient to discriminate the small $x$
resummation results.

Going from $(\o, \gamma)$ to the $(x,Q^2)$ space 
we can now show the structure function 
$F_L^{\o} (Q^{2})$ as a function of $x$ for 
different values of the hard scale $Q^2$. 
We simply used for the convolution with the soft part the 
following parametrization of the gluon 
PDF $x g(x, Q_0^2)$ at $Q_0^2=30 GeV^2$ (cf. \cite{Moch})
\begin{equation}
\label{gPDF}
x g(x, Q_0^2)=1.6\; x^{-0.3} \;(1-x)^{4.5} \;(1-0.6 \;x^{0.3})
 \end{equation}
and we have made it evolve through the DGLAP equation in the small $x$ limit at the 1 loop accuracy to obtain its $Q^2$ dependance. We then obtain the following curves.

\begin{figure}[htbp]
\begin{center}
\epsfig{file=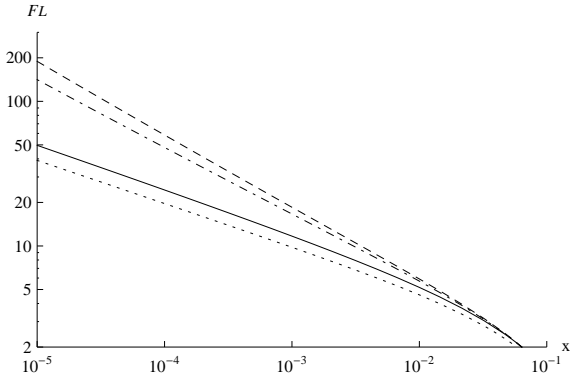,width=7.5cm}
\end{center}
\vspace{-1.0cm}
\caption{$F_L$, $Q^2 = 30 GeV^2$}
\end{figure}

\begin{figure}[htbp]
\begin{center}
\epsfig{file=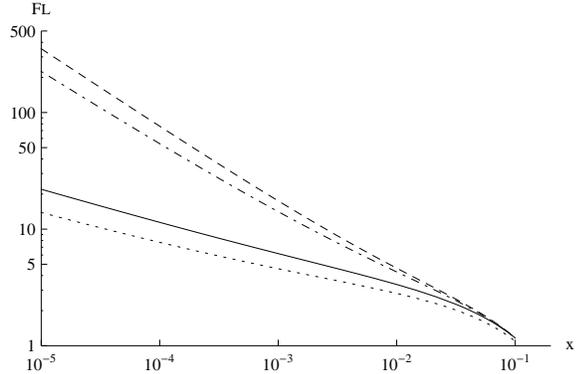,width=7.5cm}
\end{center}
\vspace{-1.0cm}
\caption{$F_L$, $Q^2 = 10 GeV^2$}
\end{figure}
 The solid curve corresponds to the Born term in 
the high energy expansion, equivalently to the 
leading order accuracy. The dotted curve contains 
the Born term and the negative 
(see eq.(\ref{numCFL})) next-to-leading order 
corrections, explaining the trend. 
Then we do the expansion of the small $x$ resummation
result in $\bar \alpha_S \ln x$ up to 12th order
which is quite sufficient because of good convergence.
This allows to show the discrepancy
for these phenomenological predictions between the 
two different K-factor expressions: 
the dashed curve corresponds to our result and 
the dashed-dotted one to the same analysis 
with the  Catani-Hautmann result. 
The convergence is very convincing and it is 
increased with the value of the hard scale.
All the curves are plotted  with the factorization 
scale $\mu^{2}_{F}=Q^2$.

\section{Exclusive electroproduction}

We now turn to the application for exlusive vector 
meson (VM) electroproduction in 
Deeply Virtual Compton Scattering (DVCS) 
following the line of  \cite{Ivanov:2007je}: 
considering the gluon dominance in the Regge limit 
of the scattering, the amplitude reads 
\bea
\label{forcalc}
{\cal I}m A^g \simeq {H^g(\xi,\xi)}
+ \int\limits^1_{2\xi}\frac{d x}{x}{H^g(x,\xi)}
\\
\times \, \sum\limits_{n=1}C^g_{VM,n}\frac{\bar \alpha_s^n}{(n-1)!}\log^{n-1}\frac{x}{\xi} \, ,\nonumber
\eea
where ${H^g(\xi,\xi)}$ is the Born contribution of 
the gluon GPD. The $C^g_{VM,n}$ are polynomials 
of  $\log\frac{Q^2}{\mu_F^2}$ obtained as in the 
previous example in the perturbative expansion of 
the coefficient function in power of 
$\gamma_{\o}^{(0)}$
\nonumber
\bea
C_{VM , \, {\o}}^g(\as , Q^{2}/\mu^{2}_{F})  \hspace{-0.2cm}&=& \hspace{-0.2cm}
h_{VM , \, {\o}} \left( \gamma_{{\o}}  \right) \;
R_{{\o}} \;(Q^{2}/\mu^{2}_{F})^{\gamma_{{\o}}} \nonumber \\
&=&  \hspace{-0.2cm}\sum\limits_{n=0} C^g_{VM,n} (\gamma_{\o}^{(0)})^n
\;\;,
\label{CLVMg}
\eea
with the following expression of the Mellin 
transform $h_{VM , \, {\o}} $ of the hard 
cross-section: we define for that the properly 
normalized impact factor $\gamma^* \to VM $ written 
as a convolution of  the hard scattering 
amplitude  with the leading twist non-perturbative 
Distribution Amplitude (DA) \cite{DA}, where both 
photon and vector meson are longitudinally polarized. 

\begin{equation}
\label{liVM}
h_{VM}(k_t^2)=\int\limits^1_0 dz \,
\frac{Q^2}{k_t^{\,\, 2}+z (1-z) Q^2}\phi_{VM} (z)
\end{equation} 
and its Mellin transform reads for an asymptotic vector meson DA $\phi_{VM} (z)= 6 z (1-z)$,
\bea
\label{galiVM}
\nonumber
 h_{VM , \, {\o}}(\gamma_{\o}) \hspace{-0.2cm}&=&\hspace{-0.2cm}\gamma_{\o} \int\limits^\infty_0\frac{d  k_t^{\,\, 2}}{k_t^{\,\, 2}}
\left(\frac{k_t^{\,\, 2}}{Q^2}\right)^{\gamma_{\o}} h_V(k_t^2)
\\
&=&\hspace{-0.2cm} \frac{\Gamma^3[1+\gamma_{\o}]\Gamma[1-\gamma_{\o}]}{\Gamma[2+2\gamma_{\o}]}\, .
\eea

We replace in the high energy term the gluon GPD by its forward limit  {$H^g(x,\xi)\to x g(x)$}. Contrarily to the previous  study for $F_L$, we keep this expression for the soft part without doing any $Q^2$-evolution. We replace the Born term by a very simple model $H^g(\xi,\xi)=1.2  \xi g(\xi)$. We obtain the following curves, in the same spirit as  previously by doing the high energy expansion.

\begin{figure}[htbp]
\begin{center}
\epsfig{file=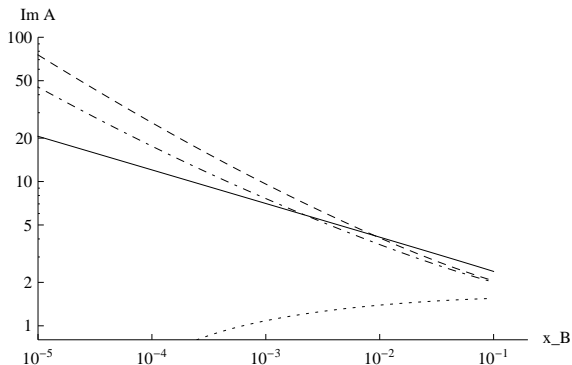,width=7.5cm}
\end{center}
\vspace{-1.0cm}
\caption{ VM electroproduction,
$Q^2 = 30 GeV^2$}
\end{figure}
The solid curve corresponds to the Born term, 
the dotted curve to   the Born term with the 
NLO corrections. 
Note that the numerical value of the $C^g_{VM,1} $ 
coefficient is much larger than in the $F_L$ case, 
giving these stronger and very negative NLO 
corrections.  This appears as an instability
in the perturbative prediction 
for this process. Here the small $x$ resummation is
really needed to obtain a reasonable and stable
prediction \cite{Ivanov:2007je}.  
After twelve iterations we get the dashed curve 
corresponding to our result and the dashed-dotted 
one corresponding to the same analysis 
with the  Catani-Hautmann K-factor. 
Also here  the convergence of the series 
expansion after twelve iterations is very good.  
Although the curves are only plotted  here with 
the factorization scale $\mu^{2}_{F}=Q^2$, 
we observe that the factorization scale dependence 
is reduced when taking into account the 
high energy resummations compared to the Born or 
even NLO case. We also note that the sensitivity 
of this choice is equivalent for both K-factor 
expressions.

\begin{figure}[htbp]
\begin{center}
\epsfig{file=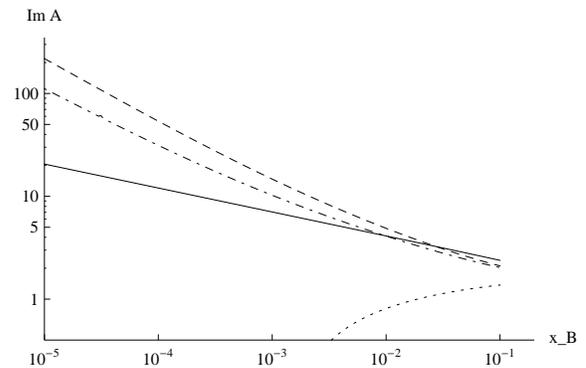,width=7.5cm}
\end{center}
\vspace{-1.0cm}
\caption{VM electroproduction,
$Q^2 = 10 GeV^2$}
\end{figure}


\section*{Acknowledgments}

We are grateful to D. Yu. Ivanov for useful 
discussions and correspondance. 
This work is  supported by  the DFG 
(contract KI-623/4).

\end{document}